\documentstyle[12pt,psfig]{article}

\def\thefootnote{\fnsymbol{footnote}}

\newcommand{\be}{\begin{equation}}	
\newcommand{\ee}{\end{equation}}	
\newcommand{\beq}{\begin{eqnarray}}	
\newcommand{\eeq}{\end{eqnarray}}	
\newcommand{\beqstar}{\begin{eqnarray*}}	
\newcommand{\eeqstar}{\end{eqnarray*}}

\newcommand{\lsim}{ \mathop{}_{\textstyle \sim}^{\textstyle <} }
\newcommand{\vev}[1]{ \left\langle {#1} \right\rangle }

\newcommand{\MG}{M_{\rm GUT}}
\newcommand{\MP}{M_{\rm Pl}}
\newcommand{\Hb}{\bar{H}}
\newcommand{\fb}{\bar{\phi}}
\newcommand{\ov}[1]{\overline{#1}}

\begin{document}

\begin{titlepage}

\begin{minipage}[t]{3in}
\begin{flushleft}
April, 1999
\end{flushleft}
\end{minipage}
\hfill
\begin{minipage}[t]{3in}
\begin{flushright}
Fermilab-PUB-99/094-T\\
hep-ph/9904252
\end{flushright}
\end{minipage}

\begin{center}
{\Large \bf Doublet-Triplet Splitting and Fermion Masses\\
with Extra Dimensions} 

\vskip .5in 

{\large Hsin-Chia Cheng}\footnote{Email hcheng@fnal.gov}

\vskip .5in
 
{\it Fermi National Accelerator Laboratory,\\
     P.O. Box 500,
     Batavia, IL 60510}

\vskip .5in

\begin{abstract}
{\normalsize  The pseudo-Goldstone boson mechanism for the
``doublet-triplet splitting'' problem of the grand unified theory can
be naturally implemented in the scenario with extra dimensions and
branes.  The two SU(6) global symmetries of the Higgs sector are
located on two separate branes while the SU(6) gauge symmetry is in
the bulk. After including several vector-like fields in the bulk, and
allowing the most general interactions with their natural strength
(including the higher dimensional ones which may be  generated by
gravity) which are consistent with the geometry, a realistic pattern
of the Standard Model fermion masses and  mixings can be naturally
obtained without any flavor symmetry.  Neutrino masses and mixings
required for the solar and  atmospheric neutrino problems can also be
accommodated.  The geometry of extra dimensions and branes provides
another way to realize the absence of certain interactions (as
required in the pseudo-Goldstone boson mechanism) or the smallness of
some couplings ({\it e.g.,} the Yukawa couplings between the fermions
and the Higgs bosons), in addition to the usual symmetry arguments.}

\end{abstract}

\vskip 1in
PACS number: 12.10.Dm 12.60.Jv 12.15.Ff 12.90.+b

\end{center}
\end{titlepage}

\renewcommand{\thefootnote}{\arabic{footnote}}
\setcounter{footnote}{0}
\pagestyle{plain} 

\section{Introduction}

The Standard Model (SM) fermions forming complete multiplets of a
single gauge group, and the unification of the SU(3)$_C$, SU(2)$_W$,
and U(1)$_Y$ gauge couplings of the Minimal Supersymmetric Standard
Model (MSSM) at $\sim 10^{16}$ GeV are strong suggestions that there
is a grand unified gauge group (SU(5) or bigger) at a very high energy
scale. However, the successful prediction of the gauge coupling
unification and the proton decay constraint require the triplet
partners of the two light Higgs doublets to have masses of the order
of the grand unification scale. The ``doublet-triplet splitting''
problem is the most problematic aspect of a grand unified theory
(GUT). There exist several solutions to this problem.  However, to get
a complete grand unified model which incorporates  these solutions in
a simple and appealing way is not easy.

One of the most appealing solutions to the ``doublet-triplet
splitting'' problem is the pseudo-Goldstone bosons (PGB)
mechanism~\cite{PGB,BDM,BCR}, where the Higgs doublets remain light
because they belong to the pseudo-Goldstone multiplets coming from
breaking of the enlarged global symmetry of the Higgs superpotential.
Here we briefly review it. The model is based on the gauge group
SU(6). The Higgs sector consists of an adjoint ({\bf 35}),  $\Sigma$,
and a pair of fundamental ({\bf 6}) and anti-fundamental ($\bar{\bf
6}$), $H$ and $\Hb$. Provided  no cross coupling exists between
$\Sigma$ and $H$, $\Hb$,  there is an effective SU(6)$_\Sigma
\times$SU(6)$_H$ symmetry of the Higgs sector. The $\Sigma$ and $H$,
$\Hb$ Higgses develop the following vacuum expectation values (vevs),
\beq
\label{vevs}
\vev{\Sigma}&=& {1\over \sqrt{12}}{\rm diag}
(1, 1, 1, 1, -2, -2) v_\Sigma ,\\
\label{vevh}
\vev{H}&=&\langle\Hb\rangle =(1, 0, 0, 0, 0, 0) v_H.
\eeq
These two SU(6)'s are then broken down to
SU(4)$\times$SU(2)$\times$U(1) and SU(5) respectively, while the SU(6)
gauge symmetry is broken down to the SM gauge group
SU(3)$_C\times$SU(2)$_\times$U(1)$_Y$.  The successful prediction of
the $\sin^2 \theta_W$ is preserved if $v_H > v_\Sigma$.  After
counting the number of the Goldstone modes and the  broken gauge
generators, one finds that there are two electroweak doublets not
eaten by the gauge bosons and hence left massless.  They are linear
combinations of the electroweak doublets coming from $\Sigma$ and $H,
\Hb$ fields,
\be
h_u={v_H H_{\Sigma} - \sqrt{3 \over 4}v_\Sigma H_H \over \sqrt{v_H^2 
+{3\over 4}v_{\Sigma}^2}}, \,\,\,
h_d={v_H \Hb_{\Sigma} - \sqrt{3\over 4} v_\Sigma \Hb_{\Hb} \over 
\sqrt{v_H^2 +{3\over 4}v_{\Sigma}^2}},
\ee
which parametrize the flat direction of the relative orientations of
the $\Sigma$ and $H$, $\Hb$ vevs. After  including the soft
supersymmetry (SUSY) breaking terms, the vevs will shift by an amount
of the order of soft SUSY breaking parameters. This shift naturally
generates a $\mu$ term, $\mu h_u h_d$, of the same order as the soft
SUSY breaking  terms from the superpotential. The Higgs potential
\be
V(h_u, h_d)= m_1^2 h_d^\dagger h_d + m_2^2 h_u^\dagger h_u
+ m_3^2 (h_u h_d + \mbox{h.c.}) + D\mbox{-terms},
\ee
(where $m_1^2=m_d^2+\mu^2$, $m_2^2=m_u^2+\mu^2$, $m_3^2=B\mu$,
$m_u^2$, $m_d^2$ are soft SUSY breaking mass terms for the
up- and down-type Higgses and $B$ is the soft SUSY breaking
parameter associated with the $\mu$-term,) satisfies the
boundary condition
\be
m_1^2(M_G)=m_2^2(M_G)=-m_3^2(M_G)
\ee
at the GUT scale $M_G$, so that there are two massless doublet bosons
corresponding to the Goldstone modes. This provide a constraint on the
phenomenology of this model~\cite{BDM,CR}. The effective
SU(6)$\times$SU(6) symmetry is explicitly broken by the couplings of
matter fields to both $\Sigma$ and $H\Hb$. The radiative corrections
from these couplings lift the flat direction and it is possible to
obtain the desired electroweak symmetry breaking Higgs potential after
running down to the low scale~\cite{BDM,CR}.

The problem of this model is that the cross coupling $H\Sigma \Hb$ is
allowed by the gauge symmetry. If it exists, it destroys the
SU(6)$_\Sigma\times$SU(6)$_H$ global symmetry of the Higgs sector and
therefore the  PGB mechanism for the light Higgs doublets. Some extra
discrete symmetries or larger gauge symmetry are needed to forbid this
coupling~\cite{BDM,BCR,Be}. Besides,  as one expects from the quantum
gravity effects, all higher dimensional operators suppressed by the
Planck scale ($\MP$), which are allowed by symmetries, might be
present and have ${\cal O}(1)$ coefficients. If that is  true, then
because $\MG/\MP$ is not a big suppression factor, the extra
symmetries have to forbid the cross couplings  between $\Sigma$ and
$H$, $\Hb$ to very high orders.  This may require some unappealing
symmetries or charge assignments.  It would be desirable to have some
better ways to suppress these unwanted couplings. As we will see in
the next section, it can be naturally achieved if there are extra
dimensions in which the gauge and  some matter fields can propagate
while $\Sigma$ and $H$, $\Hb$ are localized on two separate
branes. This provides a different mechanism to forbid the unwanted
interactions from symmetry reasons.

Another problem of this model is how to obtain the SM fermion masses.
In SU(6) models, a family of light matter fields (quarks and leptons)
can be contained in $\bf 15\, +\, \ov{6}\, +\, \ov{6}$, which is the
smallest anomaly-free combination of chiral representations. However,
there is no renormalizable Yukawa  coupling between the light fermions
belonging to $\bf 15\, +\, \ov{6}\, +\, \ov{6}$ and the light
Higgses. In order to get the large  top Yukawa coupling, one can
introduce a $\bf 20$, a pseudo-real representation, which contains the
top quark (${\bf 10}_3$ of SU(5)), then the top quark is naturally the
only one which can receive an ${\cal O}(1)$ coupling from the
interaction ${\bf 20}\,\Sigma \,{\bf 20}$. Other fermions can get
masses from the nonrenormalizable operators and therefore are
naturally suppressed. However, if all nonrenormalizable operators
consistent with the gauge symmetry exist, a realistic fermion  mass
pattern is not obtained. Therefore, one also has to introduce  extra
discrete symmetries and assume that the higher dimensional operators
are  generated by integrating out some heavy vector-like
fields~\cite{BDSBH,Be} in  order to obtain a realistic pattern of
fermion masses and mixings.  In section~\ref{fermion} we will find
that the geometry of extra dimensions and branes for the PGB mechanism
can also help to explain the fermion mass hierarchies without
appealing to flavor symmetries.

\section{Doublet-triplet splitting in extra dimensions}

Now let us discuss how the doublet-triplet splitting and fermion mass
hierarchies can naturally arise in the scenario with extra dimensions
and branes. We assume that the SU(6) gauge field propagates in the
bulk of a $4+n$ dimensional space-time with $n$ dimensions of space
compactified with a radius $R$.  The two kinds of Higgses $\Sigma$ and
$H$, $\Hb$ on the other hand  are localized on two parallel 3-branes
separated by a distance $r$ in the $4+n$ dimensions, so there is no
direct interaction between them. Extra dimensions with
compactification radius larger than the Planck (or string) length have
been considered in string theory~\cite{An,Ly,HW,Co}; they have been
used to lower the unification scale~\cite{DDG}; a very large
compactification radius can even push the fundamental Planck scale,
$M_*$, down to ${\cal O}$(TeV), providing an alternative solution to
the hierarchy problem~\cite{ADD,ADD2,ADM}. In this paper we consider the
compactification of extra dimensions occurs at high energies, around
the GUT scale, so that the successful gauge coupling unification still
works in the traditional way.\footnote{In fact, the simple SUSY GUT
prediction of the  strong coupling constant is a little higher than
the experimental value. If $1/R$ is smaller than $\MG$, the
contribution from the Kaluza-Klein states of the gauge fields will
lower the  prediction of the strong coupling constant, so it may be
favorable to have $1/R$ a little bit lower than $\MG$. We will not
discuss this in details. See the Refs.~\cite{DDG,Carone} for the
discussions of gauge coupling unification.}  We assume that the
distance between these two 3-branes, $r$, is  much smaller than the
compactification radius $R$, but larger than the fundamental Planck
distance $1/M_*$, so that we can still use the field theory
description without dealing with the full quantum gravitational
theory.\footnote{The validity of the field theory description may be
questioned at the scale very close to $M_*$. However, without knowing
how to describe the full quantum gravitational theory, we assume that
just beneath $M_*$, physics can be described by the usual field theory
with the gravity effects included in the higher dimensional
interactions suppressed by $M_*$.}
Therefore we have
\be
\MG \,\lsim \,\frac{1}{R}\, < \,\frac{1}{r}\, < \,M_* \,< \,\MP,
\ee
where $\MP \simeq 2.8\times 10^{18}$GeV is the effective
4-dimensional Planck scale. If there are no additional large extra 
dimensions in which gravitons propagate,
$\MP$ is related to $M_*$ and $R$ by~\cite{ADD,ADD2}
\be
\frac{\MP^2}{M_*^2} = M_*^n R^n .
\ee
This relation can be modified if there are additional large 
dimensions in which gravitons propagate.

On these two branes, we assume for simplicity that the Higgs 
superpotential takes the simple form,
\beq
W_1&=&m_\Sigma \Sigma^2 +\Sigma^3, \\
W_2&=& S(H\Hb -v_H^2),
\eeq
where $S$ is a singlet field, so that $\Sigma$ and $H$, $\Hb$
acquire vevs of the form given by Eqs.(\ref{vevs}),(\ref{vevh}).
To preserve the gauge coupling unification we need
\be
\MG\, = \, v_\Sigma \, < \, v_H \, (<\, M_*),
\ee
so that the light Higgses are predominantly contained in $\Sigma$.
How exactly they acquire such vevs is not important, and they may
be generated dynamically~\cite{GUTscale}.

Most of the SM matter fields as well as some additional heavy
vector-like fields live in the bulk. We assume that the extra
dimensions are compactified on an orbifold so that the unwanted  zero
modes are projected out and we can get chiral multiplets in four
dimensions. 
At low energies, the four dimensional effective Lagrangian obtained
from integrating out the extra dimensions (or equivalently from a
pure four dimensional point of view, integrating out the heavy
Kaluza-Klein towers of the bulk fields) will contain light fields
coming from both branes and the bulk.
The SU(6)$_\Sigma\times$SU(6)$_H$ global symmetry on these
two branes is broken by the couplings of the matter fields living in
the bulk to the Higgses on both branes.  Nevertheless, if we assign a
matter parity (which is  equivalent to the $R$-parity of the MSSM)
$-1$ to all fields living in the bulk, and $+1$ to the Higgses
$\Sigma$ and $H$, $\Hb$, then any interaction between the $\Sigma$
or $H$, $\Hb$ and the bulk fields must contain at least two fields
with parity $-1$.
One can easily see that any diagram having both $\Sigma$ and 
$H$, $\Hb$ as only external lines contains a loop at least.
By the non-renormalization theorem, no direct superpotential couplings
between $\Sigma$ and $H$, $\Hb$ (and containing no matter fields) at
any order can be generated after integrating out the extra
dimensions. 
Thus, the PGB mechanism for the doublet-triplet splitting
can work naturally in this scenario.

\section{Fermion masses}
\label{fermion}

Before getting into the details of the fermion masses and mixings in
the Standard Model, we first discuss in general the possible
suppressions of couplings we may get in such a scenario. In addition
to the usual Planck mass suppression for the higher dimensional
operators, the suppressions may also come from the large volume factor
of the extra dimensions and from integrating out the vector-like bulk
fields and their  Kaluza-Klein excitations.

The couplings of an (external) bulk field (after integrating
out the extra dimensions and heavy fields) to the brane fields
are suppressed by the volume factor of the extra dimensions.
The zero modes contain an $R^{-n/2}$ factor after the Fourier
decomposition to match the mass dimensions of fields in different
space dimensions, so the dimensionless coefficients of the couplings 
are naturally suppressed by~\cite{ADD2,AD}
\be
\epsilon \equiv (M_* R)^{-\frac{n}{2}} \left( = \frac{M_*}{\MP},
\; \mbox{if no additinal large dimensions for gravitons}\right).
\ee
This may explain the weakness of the unified gauge coupling at the GUT
scale. To get ${\cal O}(1)$ Yukawa coupling for the top quark, we
therefore assume that the $\bf 20$ (denoted by $\eta$, with matter
parity $-1$) containing the top quark lives on the same brane in which
$\Sigma$ resides, then the $\eta\, \Sigma\, \eta$ interaction which
contains the top Yukawa coupling is naturally ${\cal O}(1)$.  All
other matter and vector-like fields are assumed to  live in the bulk.

Light fermion masses come from higher dimensional operators.  Higher
dimensional operators can already be present in the fundamental
Lagrangian (suppressed by powers of $M_*$) if they involve fields in
the bulk and on one brane only. They can also be generated by
integrating out the heavy vector-like fields in the bulk and extra
dimensions if they contain fields on both branes. This is  somewhat
similar to the Froggatt-Nielsen mechanism~\cite{FN}.  However, the
suppression of these higher dimensional operators is different. It
also depends on the transverse distance $r$ between the two branes and
the number of extra  dimensions, as there is a tower of the
Kaluza-Klein states of the vector-like fields. The case when there are
vector-like  scalars connecting two branes is discussed in
Ref.~\cite{AD}.  It is simply the Yukawa potential (or the propagator
of the vector-like field) in the $n$ transverse direction. The
generalization to the supersymmetric case is straightforward.  The
propagator in the transverse direction is
\be
\Delta_V (r)= \int d^n \kappa \; e^{i\kappa r} {-i\kappa+m_V
\over \kappa^2+m_V^2}.
\ee
One gets an exponential suppression ($e^{-m_V r}$) if the mass of the
vector-like fields $m_V$ is larger than $1/r$, and a power suppression
($r^{1-n}$) if $m_V$ is smaller than $1/r$. For one extra dimension
and $m_V < 1/r$,  there can even be no suppression.  We will
parametrize the dimensionless suppression factor (in the unit of
$M_*$) by $\delta_V$ ({\it e.g.}, $(M_* r)^{-a}$ in the case of power
suppression). We will find that to obtain  successful fermion masses
some of the  suppression factor from integrating out the vector-like
fields should be ${\cal O}(1)$, so it is favorable to have just one
extra dimension.

In the bulk, there are three sets of
${\bf 15\, +\, \ov{6}\, +\, \ov{6}}$ chiral matter
multiplets, denoted by $\psi_i({\bf 15}),\; \fb_i({\bar{\bf 6}}),\;
\fb'_i({\bar{\bf 6}}),\; i=1, 2, 3$. In addition, we assume
that there are 3 pairs of vector-like fields of the SU(6)
representations $({\bf 20}_1,\, {\bf 20}_2)$, ${\bf (6,\, \ov{6})}$, 
and ${\bf (70,\, \ov{70})}$
with  masses (of the zero modes) $m_{20}$, $m_6$, and $m_{70}$
respectively. They all have matter parity $-1$. The field content
is summarized in Table~\ref{content}.
\begin{table}[htb]
\begin{center}
\begin{tabular}{|c|c|c|} \hline
Brane 1 & Bulk & Brane 2 \\ \hline
$\Sigma$ & SU(6) gauge field,
$\psi_i,\, \bar{\phi}_i, \, \bar{\phi}'_i, \, i=1, 2, 3$ &
$H, \, \Hb$ \\ 
$\eta$ & ${\bf 20}_1, {\bf 20}_2, \, {\bf 6, \ov{6}, \, 70, 
\ov{70}}$ & $S, \; (N)$ \\
\hline
\end{tabular}
\end{center}
\caption{Field content in the bulk and on the two branes.
As it will be discussed later, the singlet field $N$ is included 
if we need to generate the neutrino mass to account for the 
atmospheric neutrino oscillation.
\label{content}}
\end{table}
In terms of the usual SU(5)$_{\rm GUT}$ subgroup, these
representations decompose into:
\beq
{\bf 6} &=& {\bf 1+5, \quad \ov{6}=1+\ov{5},} \nonumber \\
{\bf 15} &=& {\bf 5+10,} \nonumber \\
{\bf 20} &=& {\bf 10 + \ov{10},} \nonumber \\
{\bf 35} &=& {\bf 1+5+\ov{5}+24,} \nonumber \\
{\bf 70} &=& {\bf 5+10+15+\ov{40}, \quad \ov{70}=\ov{5}+\ov{10}+\ov{15}+40,}
\eeq

Integrating out the vector-like fields and the extra dimensions, we
obtain the operators appearing in the low energy effective four dimensional
theory. The dimensionless coefficient (after factorizing out powers of
$M_*$ of the dimensionful coupling) of an operator  is suppressed by a
power of $\epsilon$ for  each external bulk field, and by $\delta_V$
if it is generated by integrating out the vector-like fields $V,
\ov{V}$.  For Yukawa couplings coming from nonrenormalizable
operators, they will also be suppressed by $v_H/M_*$ or
$v_\Sigma/M_*$. If the light Higgs doublets come from $H$, $\Hb$,
there is a further suppression of the mixing angle $\sim v_\Sigma
/v_H$. In the following we discuss these effective operators and the
SM fermion masses and mixings arising from them. For simplicity and
the organization purpose, the operators are written in terms of
the SU(6) language. What we really mean are the operators involving
the light fields contained in those operators, since the heavy fields 
($\sim M_{\rm GUT}$) in these operators should
have been integrated out too.
For example,
$\eta \Sigma \psi_2 H$ means ${\bf 10}_{\eta} {h_u}_{\Sigma}
{\bf 10}_{\psi_2} \langle H \rangle$ (and the higher order
term ${\bf 10}_{\eta} 
\langle \Sigma \rangle {\bf 10}_{\psi_2} {h_u}_ H$).

\begin{description}
\item [Operators which decouple the extra states] (Fig.~\ref{heavy}):
\begin{figure}[htbp]
\centerline{\psfig{file=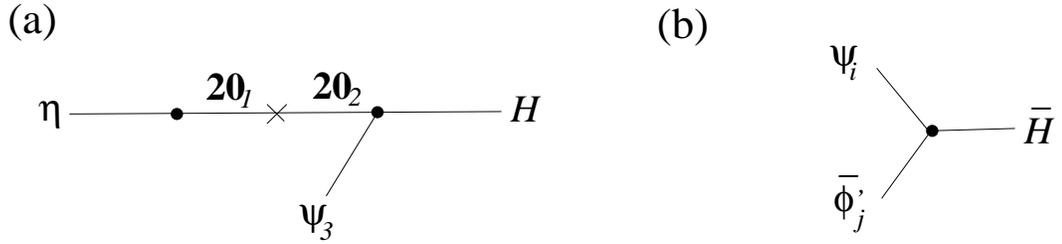,width=0.9\textwidth}}
\caption{Diagrams which decouple the extra states. Fields on the
left of the left interaction point are on brane 1. Fields on the
right of the right interaction point are on brane 2. Fields between
the interaction points are in the bulk. }
\label{heavy}
\end{figure}
\begin{itemize}
\item 
$\eta H \psi_3$ (diagram (a)): We can define $\psi_3$
with this operator by using the rotation freedom among $\psi_i$'s,
and ${\bf 20}_1$ to be the one which couples to $\eta$. The ${\bf 10}$
(of SU(5)$_{\rm GUT}$) in $\psi_3$ and $\ov{\bf 10}$ in $\eta$
become heavy due to $\langle H \rangle = v_H$, leaving only
three {\bf 10}'s (in $\eta, \, \psi_2 , \, \psi_1$) in the low energies.
\item
$\psi_i \Hb \bar{\phi}'_j$ (diagram (b)): The {\bf 5}'s in
$\psi_i$ and $\ov{\bf 5}$'s in $\bar{\phi}'_j$ are married by
$\langle \Hb \rangle$, leaving only three $\ov{\bf 5}$'s (in 
$\bar{\phi}_i$) in the low energies.
\end{itemize}
Because of the suppresion factors involved, some decoupled states
will have masses a little bit lower than $\MG$. However, they are
complete SU(5) multiplets, so they do not affect the coupling
unification.  We can see that $\psi_3$ is completely decoupled, so we
will drop it in the following discussion. In SU(5) notation, the three
light generations are contained in the ${\bf 10}$'s of $\eta, \,
\psi_2, \, \psi_1$, and $\ov{\bf 5}$'s of $\bar{\phi}_3, \,
\bar{\phi}_2, \, \bar{\phi}_1$. They are the only SM non-singlets
matter fields left massless at this stage. (The SU(5) singlets can
also be decoupled. It will be seen when we discuss neutrino masses.)

\item [Up-type quark masses] (Fig.~\ref{up}):
\begin{figure}[htbp]
\centerline{\psfig{file=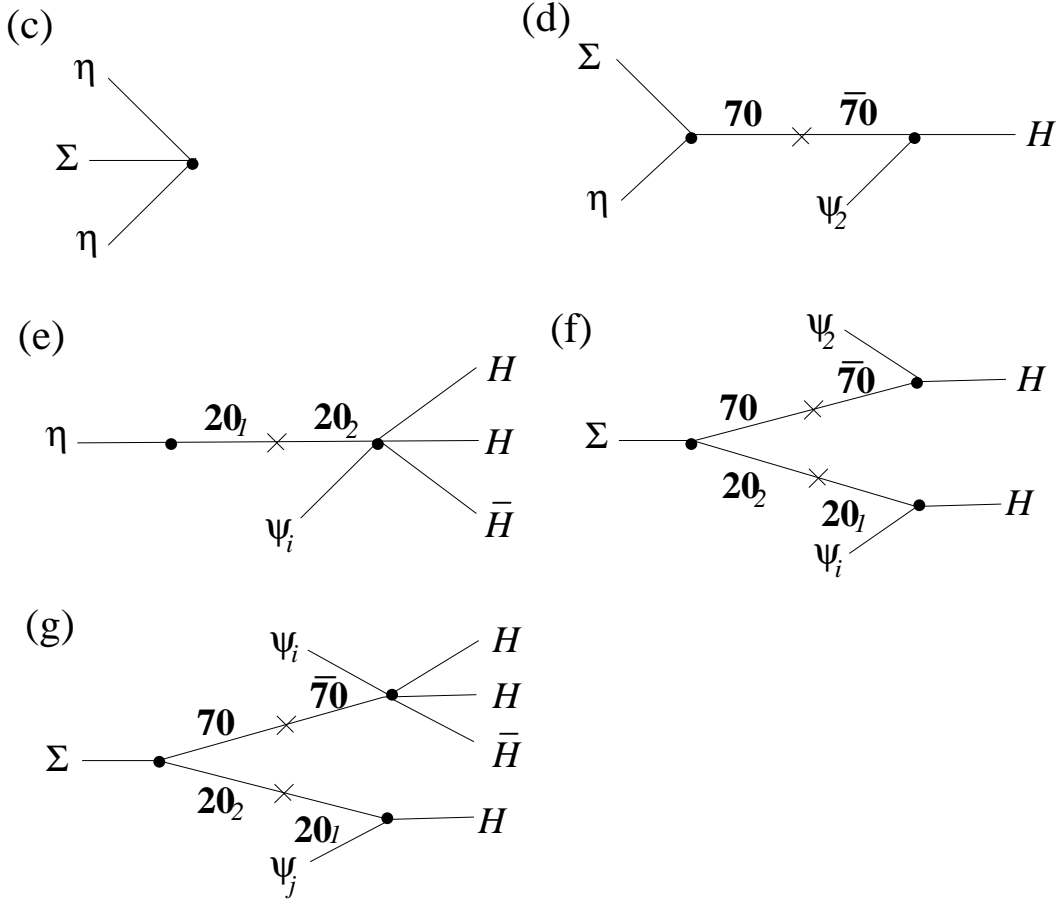,width=0.9\textwidth}}
\caption{Diagrams which generate the up-type quark masses.}
\label{up}
\end{figure}
\begin{itemize}
\item
$\eta \Sigma \eta$ (diagram (c)): It contains ${\bf 10}_3
{\bf 5}_{\Sigma} {\bf 10}_3$ in SU(5) notation. 
There is no suppression and therefore
it gives an ${\cal O}(1)$ Yukawa coupling to the top quark.
\item
$\eta \Sigma \psi_2 H$ (diagram (d)): One can rotate $\psi_i,\;
i=1,2$, so that only $\psi_2$ couples to $\ov{\bf 70}$ and $H$.
It generates the 23 and 32 elements of the up Yukawa matrix
of ${\cal O}(\epsilon \delta_{70} ({v_H \over M_*}))$.
\item
$\eta \psi_i H (H \Hb)$ (diagram (e)): It does not contain $\Sigma$,
so the light Higgs has to come from $H$, which causes a
$({v_{\Sigma}\over v_H})$ mixing suppression. It generates 13, 31, 23, 32
elements of ${\cal O}(\epsilon \delta_{20} ({v_H\over M_*})^2
({v_{\Sigma} \over v_H}))$.
\item
$\Sigma \psi_2 \psi_i H H$ (diagram (f)): It generates 22, 12, 21
elements of ${\cal O}(\epsilon^2 \delta_{20} \delta_{70}
({v_H \over M_*})^2)$.
\item
$\Sigma \psi_i \psi_j H H (H \Hb)$ (diagram (g)): One can always
attach a pair of $(H \Hb)$ to the interaction on the brane 2,
which will be suppressed by an extra $(v_H/M_*)^2$. This gives
the leading contribution to the 11 element.
\end{itemize}
In the leading order the up-type Yukawa matrix looks like
\be
\lambda_U \sim \left( \begin{array}{ccc}
                       u_4 & u'_3 & u_2 \\
                       u'_3 & u_3 & u_1 \\
                       u_2 & u_1 & 1 
                      \end{array} \right),
\ee
where 
\beq
u_1 &\sim & \epsilon \delta_{70} \left({v_H \over M_*}\right),\\
u_2 &\sim & \epsilon \delta_{20} \left({v_H \over M_*}\right)^2 
\left({v_{\Sigma} \over v_H}\right),\\
u_3,\, u'_3 &\sim & \epsilon^2 \delta_{20} \delta_{70}
\left({v_H \over M_*}\right)^2,\\
u_4 &\sim & \epsilon^2 \delta_{20} \delta_{70}
\left({v_H \over M_*}\right)^4.
\eeq
If we take $\epsilon \sim {1\over 3}$, ${v_H \over M_*} \sim
{1\over 5}$, ${v_{\Sigma} \over v_H} \sim {1\over 3}$,
$\delta_{70} \sim {1\over 2}$, and $\delta_{20} \sim {1\over 40}$,
then we have at the GUT scale,
\beq
\lambda_t &\sim & 1 ,\\
\lambda_c &\sim & u_1^2 \sim \epsilon^2 \delta_{70}^2 
\left({v_H \over M_*}\right)^2 \sim 10^{-3} ,\\
\lambda_u &\sim & u_4, \, {{u'}_3^2\over u_1^2} \,
\mbox{(two comparable contributions)} \sim (2-3)\times 10^{-6} .
\eeq
Remember that the light fermion Yukawa couplings will increase
in renormalization group (RG) running to low energies while the
top Yukawa coupling will roughly approach some fixed point.
The mass ratios of light quarks to the top quark will be enhanced
by a factor of 5--10 relative to those at the GUT scale.
After taking into account the RG effect, the above numbers give
a good approximation to the up-type quark masses.
In diagonalizing the mass matrix, the 23 rotation angle $U_{23}
\sim u_1 \sim 3\times 10^{-2}$ is about the same order as
$V_{cb}$. Other rotation angles are much smaller than the
corresponding Cabibbio-Kobayashi-Maskawa (CKM) matrix elements,
so they have to be generated from the down sector.

\item [Down-type quark and charged lepton masses] (Fig.~\ref{down}):
\begin{figure}[htbp]
\centerline{\psfig{file=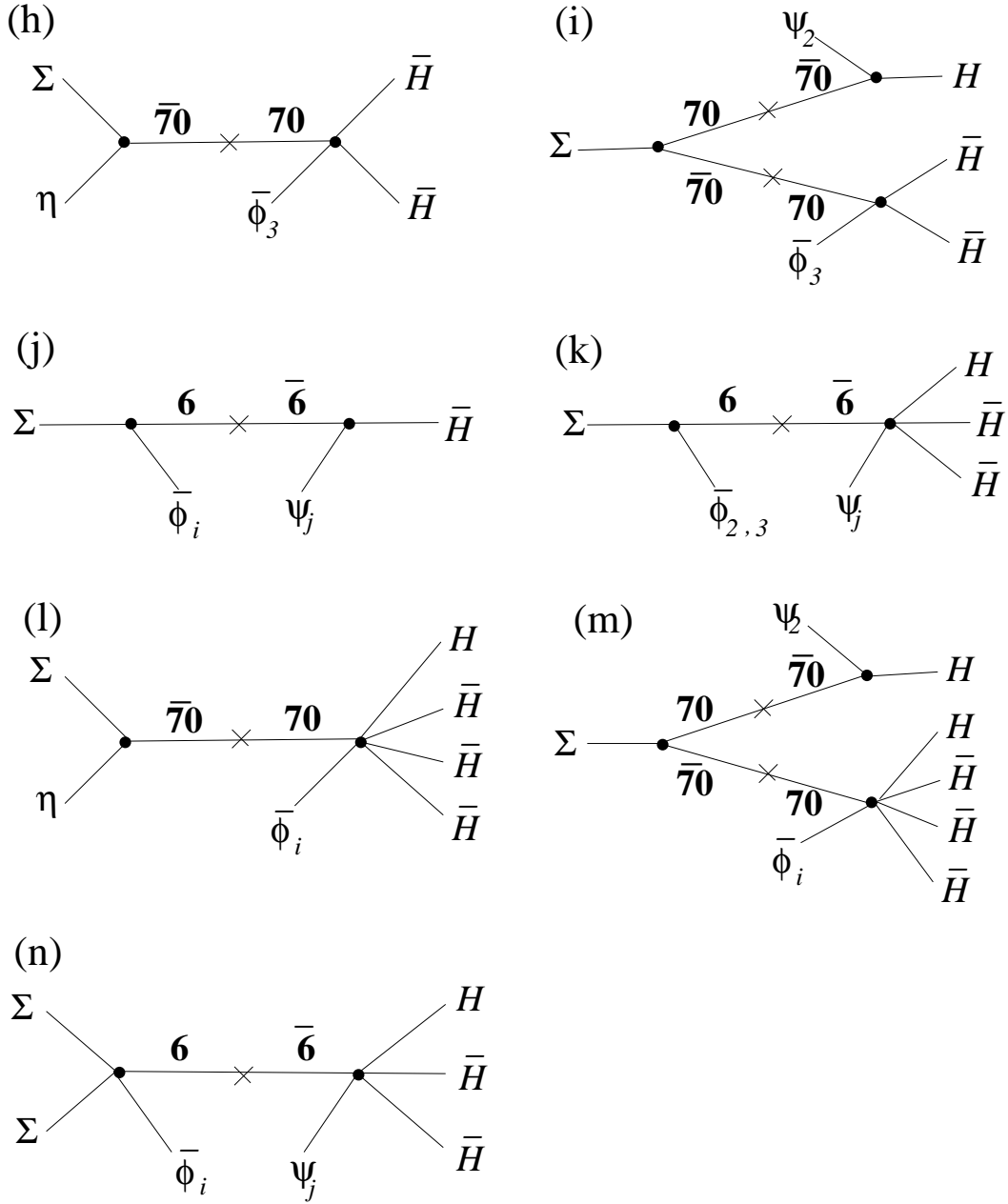,width=0.9\textwidth}}
\caption{Diagrams which generate the down-type quark and 
charged lepton masses.}
\label{down}
\end{figure}
\begin{itemize}
\item
$\eta \Sigma \bar{\phi}_3 \Hb \Hb$ (diagram (h)): We can use
the rotation freedom among $\bar{\phi}_i$'s to define $\bar{\phi}_3$
with this operator. This gives the 33 elements of the down and charged
lepton Yukawa matrix. In leading order, the Higgs doublet to which the
fermions couple comes from $\Sigma$. They are ${\cal O}(\epsilon
\delta_{70} ({v_H \over M_*})^2)$ and the same for the down-type 
quark and the charged lepton, so we have appproximate $b-\tau$
unification.
\item
$\Sigma \psi_2 H \bar{\phi}_3 \Hb \Hb$ (diagram (i)): $\psi_2$
and $\bar{\phi}_3$ have been defined before. This operator
contributes to the 23 elements of the down and lepton Yukawa
matrices and is ${\cal O}(\epsilon^2 \delta_{70}^2
({v_H\over M_*})^3)$.
\item
$\langle\Sigma\rangle \bar{\phi}_i \psi_j \Hb$ (diagram (j)):
This operator only redefines $\bar{\phi}'_i$ and is irrelevant
for fermion masses~\cite{BDSBH}.
\item
$\langle\Sigma\rangle \bar{\phi}_i \psi_j \Hb (H \Hb)$ (diagram (k)):
Attaching $(H \Hb)$ to the previous diagram, we can get a diagram
contributing to the fermion masses. We can rotate $\bar{\phi}_i$
to have only $\bar{\phi}_2,\, \bar{\phi}_3$ in this operator.
If we had not defined $\psi_i$'s in the up sector, we could also
have defined $\psi_2$ by this operator, then it would have contributed
only to the 22 and 23 elements of the mass matrices. The rotation
angle between the two bases will in general be ${\cal O}(1)$,
which accounts for why the Cabibbo angle is large. In the basis
used for the up sector, it contributes to the 12, 13, 22, 23
elements of the down and charged lepton mass matrices (with
12, 13 elements smaller than 22, 23 elements by $\sim \sin\theta_C
\sim 0.2$). An important fact of this operator is that the
intermediate states $({\bf 6},\, \ov{\bf 6})$ do not contain
${\bf 10}, \, \ov{\bf 10}$ of SU(5)$_{\rm GUT}$, so the light Higgs
doublet has to come from $\Hb$. The contribution of this
operator to the Yukawa couplings is therefore ${\cal O}
(\epsilon^2 \delta_6 ({v_{\Sigma}\over M_*})({v_H \over M_*})^2
({v_{\Sigma}\over v_H}))$. The vev of $\Sigma$ gives a ratio of
$1 : -2$ to the down-type quark and the charged lepton Yukawa
matrix elements. Since it is the dominant term
to the 22 elements and hence the leading contribution to the
second generation masses, this offers an explanation of the
discrepancy between $m_s$ and $m_{\mu}$ from the simple unification
relation.
\end{itemize}
Other matrix elements and non-leading contributions can be
obtained by attaching more $(H \Hb)$ or $\Sigma$ to previous
diagrams. In the following we only discuss the leading 
contributions.
\begin{itemize}
\item
$\eta \Sigma \bar{\phi}_i \Hb \Hb (H \Hb)$ (diagram (l)):
This gives the leading contribution to the 31, 32 elements of 
the down and lepton mass matrices of ${\cal O}(\epsilon
\delta_{70} ({v_H \over M_*})^4)$.
\item
$\Sigma \psi_2 H \bar{\phi}_i \Hb \Hb (H \Hb)$ (diagram (m)):
This gives the leading contribution to the
21 elements of the down and lepton mass matrices of
${\cal O}(\epsilon^2 \delta_{70}^2 ({v_H \over M_*})^5)$.
\item
$\langle\Sigma\rangle^2 \bar{\phi}_i \psi_j \Hb (H \Hb)$ (diagram (n)):
This gives the leading contribution to the 11 elements of
${\cal O}(\epsilon^2 \delta_6 ({v_{\Sigma}\over M_*})^2
({v_H \over M_*})^2 ({v_{\Sigma}\over v_H}))$.
\end{itemize}
In the leading order the down-type Yukawa matrix looks like
\be
\lambda_D \sim \left( \begin{array}{ccl}
                       d_6 & sd_3 & sd'_3 \\
                       d_5 & d_3 & d_2(+d'_3) \\
                       d_4 & d_4 & d_1 
                      \end{array} \right),
\ee
where
\beq
d_1 &\sim & \epsilon \delta_{70} \left({v_H \over M_*}\right)^2, \\
d_2 &\sim & \epsilon^2 \delta_{70}^2 \left({v_H \over M_*}\right)^3, \\
d_3,\, d'_3 &\sim & \epsilon^2 \delta_{6} 
\left({v_{\Sigma} \over v_H}\right)^2 \left({v_H \over M_*}\right)^3, \\
d_4 &\sim & \epsilon \delta_{70} \left({v_H \over M_*}\right)^4, \\
d_5 &\sim & \epsilon^2 \delta_{70}^2 \left({v_H \over M_*}\right)^5, \\
d_6 &\sim & \epsilon^2 \delta_{6} \left({v_{\Sigma} \over v_H}\right)^3 
\left({v_H \over M_*}\right)^4,
\eeq
and we have explicitly put in the Cabibbo angle $s\sim 0.2$.
Again, taking the previous assumed suppression factors,
$\epsilon \sim {1\over 3}$, ${v_H \over M_*} \sim
{1\over 5}$, ${v_{\Sigma} \over v_H} \sim {1\over 3}$,
$\delta_{70} \sim {1\over 2}$, $\delta_{20} \sim {1\over 40}$,
with $\delta_6 \sim 1$, we have the following relations at
the GUT scale,
\beq
{\lambda_b \over \lambda_t} &\sim & \epsilon \delta_{70}
\left({v_H \over M_*}\right)^2 \sim 10^{-2}, \\
{\lambda_s \over \lambda_b} &\sim & {d_3\over d_1} \sim
\epsilon {\delta_6 \over \delta_{70}} \left({v_{\Sigma}
\over v_H}\right)^2 \left({v_H\over M_*}\right) \sim 
1.5\times 10^{-2}.
\eeq
In running down to low energies, we get enhancements of
$\sim 3-5$ for ${\lambda_b(m_b)\over \lambda_t(m_t)}$
and $\sim 2$ for ${\lambda_s(1{\rm GeV})\over \lambda_b(m_b)}$.
There are several comparable leading contributions to $\lambda_d$
after diagonalization, {\it e.g.,} $d_6,\, {(sd_3 d_2 d_4)/ (d_1 d_3)},
\, {(sd_3 d_1 d_5)/ (d_1 d_3)}$. In terms of ${\lambda_d \over
\lambda_s}$, they are
\beq
{d_6 \over d_3} &\sim & \left({v_{\Sigma}\over v_H}\right)
\left({v_H\over M_*}\right) \sim {1\over 15}, \\
{sd_3 d_2 d_4\over d_1 d_3^2} &\sim & {s \delta_{70}^2 
\left({v_H\over M_*}\right)^2 \over \delta_6 \left(
{v_{\Sigma} \over v_H}\right)^2} \sim 2\times 10^{-2}, \\
{sd_3 d_1 d_5\over d_1 d_3^2} &\sim & {s \delta_{70}^2 
\left({v_H\over M_*}\right)^2 \over \delta_6 \left(
{v_{\Sigma} \over v_H}\right)^2} \sim 2\times 10^{-2}.
\eeq
The smallness of $\lambda_e$ may due to a mild cancellation
among these contributions. We can see that we also get a very
good approximation of the down quark and charged lepton masses
(for small $\tan \beta$).

The rotation angles for diagonalizing the down quark mass
matrix are
\beq
D_{23} &\sim & {d_2\over d_1} \sim \epsilon \delta_{70}
\left({v_H\over M_*}\right) \sim 3\times 10^{-2}, \\
D_{13} &\sim & {sd'_3 \over d_1} \sim s \epsilon
{\delta_6 \over \delta_{70}} \left({v_{\Sigma} \over v_H}\right)^2
\left({v_H\over M_*}\right) \sim 3\times 10^{-3}, \\
D_{12} &\sim & s \sim 0.2 \;({\cal O}(1)).
\eeq
$U_{23}$ and $D_{23}$ are comparable and their combination gives
$V_{cb}$. Other `CKM matrix elements are dominated by the rotation
of the down sector and they are all generated at the right
magnitudes. It is quite remarkable that without any extra flavor
symmetry and allowing most general operators, the SM fermion
masses and mixings pattern is naturally obtained provided the various
mass scales are such that they produce the appropriate suppression
factors.

\item [Neutrino masses] (Fig.~\ref{neutrino}):
\begin{figure}[htbp]
\centerline{\psfig{file=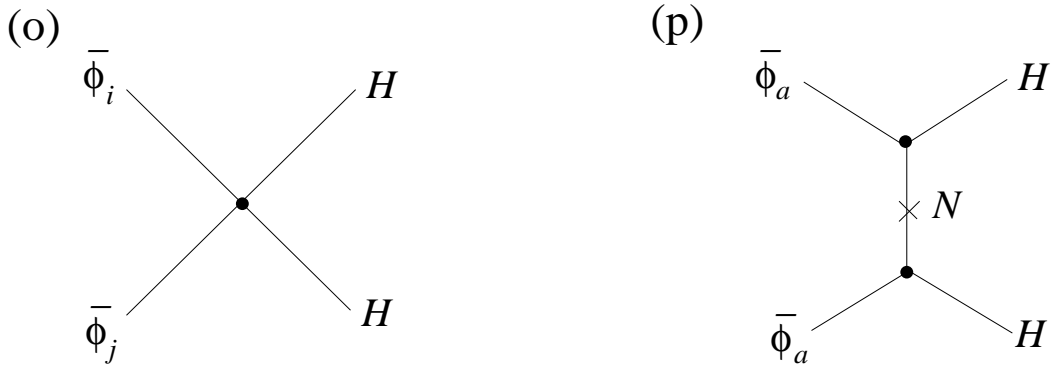,width=0.9\textwidth}}
\caption{Diagrams which generate the neutrino masses.}
\label{neutrino}
\end{figure}

Majorana masses of the left handed neutrinos can be generated by
diagram (o). This diagram also decouples the SU(5) singlets in
$\bar{\phi}_i$ (and also in $\bar{\phi}'_i$ by replacing
$\bar{\phi}_i$ by $\bar{\phi}'_i$). Because there is no distinction
among the three generations of $\bar{\phi}_i$, in general we have
large mixings among the neutrinos. The neutrino masses generated by
this diagram are of the order $\sim \epsilon^2 ({v_{\Sigma} \over
v_H})^2 {v_{\rm EW}^2\over M_*} \sim 10^{-5}-10^{-6}$~eV.  This is in
the right range of explaining the solar neutrino problem through the
``just-so'' vacuum oscillation solution~\cite{justso}, but too small
to account for the atmospheric neutrino problem, which requires
$\delta m_{\rm atm} \sim 3\times 10^{-2} -
10^{-1}$~\cite{atm,kamio}. To accommodate a larger neutrino mass,  one
can introduce a singlet field $N$ (with matter parity $-1$)  on brane
2 (which contains $H\Hb$). Then, from diagram (p), one neutrino mass
of the order $\sim \epsilon^2 ({v_{\Sigma} \over v_H})^2 {v_{\rm
EW}^2\over M_N}$ can be generated.  It will be in the right range for
the atmospheric neutrino problem if the mass of the singlet mass $M_N$
is $\sim 10^{13}-10^{14}$~GeV. The next neutrino mass obtained from
attaching $(H\Hb)$'s to diagram (p) will be suppressed by $({v_H\over
M_*})^4 \sim 10^{-3}$, close to that required for the vacuum
oscillation solution of the solar neutrino problem.

\end{description}

\section{Conclusions}

In conclusion, extra dimensions and fields localized on branes provide
a new way to understand the absence or smallness of some couplings
without symmetry arguments~\cite{AD,ADDM,RS,AS}.  This kind of idea
has been used to obtain small fermion masses in the Standard Model and
to suppress proton decay~\cite{AS}.  Here we find that by localizing
two kinds of Higgses on two separate branes, the most difficult
``doublet-triplet splitting'' problem of the grand unified theory is
naturally solved by the pseudo-Goldstone boson mechanism. In addition,
after including several vector-like fields in the bulk, and allowing
the most general interactions consistent with  the background geometry
and with their natural strength, all Standard Model fermion masses and
mixings can be correctly produced without any flavor symmetry. The
neutrino masses and mixings required for the solar and atmospheric
neutrino problems can also be easily accommodated. It is very
interesting that the complicated picture of the Standard Model can be
realized by such a simple model. Extra dimensions at such high
energies will not give us the exciting new collider signatures such as
production of the graviton Kaluza-Klein states.  Nevertheless, it
gives a simple realization of the grand unified theory and the fermion
masses with the pseudo-Goldstone boson solution to the
``doublet-triplet splitting'' problem.   If it is true, the boundary
condition of the Higgs parameters should be verified in the future
experiments.


{\bf Acknowledgements} 
The author would like to thank N. Arkani-Hamed and B.A. Dobrescu
for discussion.
Fermilab is operated by Universities Research Association, Inc.,
under contract DE-AC02-76CH03000 with U.S. Department of Energy.


%
%
\newcommand{\Journal}[4]{{#1} {\bf #2} {(#3)} {#4}}
\newcommand{\APJ}{Ap. J.}
\newcommand{\CJP}{Can. J. Phys.}
\newcommand{\NC}{Nuovo Cimento}
\newcommand{\NP}{Nucl. Phys.}
\newcommand{\MPL}{Mod. Phys. Lett.}
\newcommand{\PL}{Phys. Lett.}
\newcommand{\PR}{Phys. Rev.}
\newcommand{\PRep}{Phys. Rep.}
\newcommand{\PRL}{Phys. Rev. Lett.}
\newcommand{\PTP}{Prog. Theor. Phys.}
\newcommand{\SJNP}{Sov. J. Nucl. Phys.}
\newcommand{\ZP}{Z. Phys.}


\end{document}